# Characterizing MRO in atomistic models of vitreous SiO$_2$ generated using *ab-initio* molecular dynamics

**Sruti Sangeeta Jena, Shakti Singh, Sharat Chandra***

Materials Science Group, Homi Bhabha National Institute, Indira Gandhi Centre for Atomic Research, Kalpakkam – 603102, India

Corresponding author: Sharat Chandra

Corresponding author's e-mail: sharat@igcar.gov.in, sharat.c@gmail.com

**Abstract**

Vitreous silica is the most versatile material for scientific and commercial applications. Although large-scale atomistic models of vitreous-SiO$_2$ (v-SiO$_2$) having medium-range order (MRO) have been successfully developed by melt-quench through classical molecular dynamics, the MRO is not well studied for the smaller-scale models developed by melt-quench using *ab-initio* molecular dynamics (AIMD). In this study, we obtain atomistic models of v-SiO$_2$ by performing melt-quench simulation using AIMD. The final structure is compared with the experimental data and some recent atomistic models, on the basis of the structural properties. Since AIMD allows for the estimation of electronic structure, a detailed study of electronic properties is also done. It shows the presence of defect states mainly due to dangling bonds in the band-gap region of electronic density of states, whereas the edge-shared type of defective structures in the glassy models are found to contribute mainly in the valence band. In addition, Oxygen and Silicon vacancies as well as bridging Oxygen type of defects were created and their contributions to the band-gap were studied.

# 1. Introduction:

Vitreous silica has proved to be an extremely versatile material for a wide range of applications due to its excellent thermal properties, optical transmission along with good electrical and corrosion performance. It is a simple binary model system that gives us an insight into various features of amorphous materials. Thus, amorphous $SiO_2$ (a-$SiO_2$) is studied extensively in the field of material science, condensed matter physics, engineering, and chemistry [1]. Over the last three decades it has grabbed the interest of various researchers in both experimental [2,3] and theoretical [4-8] fields. Ever since Zachariasen [9] has given a comprehension of the atomic arrangement in glasses, there has been an urge to model atomic-scale structures of glasses. With the help of these atomic structures, we would be able to get an insight into various phenomena in glassy structures such as chemical durability, glass-forming ability, structural stability, and other properties which are difficult to determine experimentally in a disordered system as compared to their crystalline counterparts. So, computational modelling of the atomistic structure of glasses is immensely important for understanding the fundamental physics involved at the atomic level [10,11]. Various computational techniques used for this purpose are classical molecular dynamics (CMD) [12], *ab-initio* molecular dynamics (AIMD) [13,14], Monte-Carlo (MC) methods [15,16], and hybrid methods [17,18].

In this study, melt-quench simulation [10] has been used to develop glass structures. For the purpose of glass modelling, although CMD yields large-scale structures but drawbacks like employing inter-atomic potentials developed empirically and infeasibility in the calculation of electronic and magnetic properties, severely limits the use of the developed structures for further studies [19,20]. However, these shortcomings can be overcome by coupling the molecular dynamics with density functional theory (DFT). This approach is known as AIMD [13]. The cost of going *ab-initio* in glass calculations results in smaller-scale structures that can be developed within reasonable computation time [21]. AIMD has been used increasingly for modelling glassy materials. In the recent decades, research has revealed that this method has been successful in studying binary and ternary glasses [22,23]. Recent advances in supercomputers have made it possible for this approach to structurally simulate multicomponent glasses as well [24]. It gives an insight into their atomic-level structure, properties, and behaviour [25,26]. The capability of first-principles method to accurately estimate fundamental properties like electronic structure including band gaps, and optical properties allows for a deeper insight into the physics of disordered solids [27]. In addition, AIMD has the advantage of tracking the evolution of the atomic arrangement of a structure by visualizing the system at different temperatures and simulation times thereby aiding in the phase transition studies [28]. Apart from this, it is a powerful tool for calculating diffusion and transport properties [29], and the behaviour of defects in the glassy materials [27]. Thus, AIMD has helped in bridging the gap between the theoretical predictions and the experimental observations thereby enabling designing and engineering of the glassy materials with enhanced properties. The current study attempts to develop silica glass models by performing melt-quench simulation starting from 2×2×2 supercell of β-cristobalite structure, using AIMD. Although silica glass has been studied using *ab-initio* techniques like DFT earlier [30,31], reproduction of medium range order in such small structures developed using AIMD, has not been addressed earlier in the authors' knowledge.

There has been an increased interest in reproducing the MRO in glasses and finding the cause of its structural origin [32-34]. In the authors' knowledge, not many studies that characterise the MRO in silica glass models developed from melt-quench simulation in AIMD have been published till now. We report here the structural properties like the calculated neutron structure factor, First Sharp Diffraction Peak (FSDP) [35-37] and rings distribution of the developed models, which are important from the point of view of MRO. Furthermore, to characterize the glass beyond the medium range (>10 Å), properties like void distribution and density of the structure are studied.

Apart from this, a thorough study of the possible structural defects is carried out which includes defects like dangling bonds, bridging Oxygen, and Silicon and Oxygen vacancies. The effects of these defect types on the electronic properties of the structure are studied by analysing their respective electronic density of states (DOS) plots. Furthermore, structural defects could also help in understanding cracks as well as crack propagation which can be analysed as done in the references [38,39].

In modelling glasses, melt-quench simulation has the advantage that it mimics actual experimental process of glass manufacturing, albeit at higher quench rates of the order of $10^{14}$-$10^{15}$ K/sec [40,41] as compared to those seen in the experiments. The reason behind adhering to such quench rates in AIMD simulations is that we need to complete the dynamics within a very short time period of a few tens of nanoseconds, resulting in very large computational cost. As a result, quench rates of the order $10^{14}$ K/sec [42] have to be used to observe glass formation while any slower quench rate show significant crystallization in the system [43,44]. Hence, we explore the effects of faster quench rate in the present study.

Outline of rest of the paper is as follows: Section 2 describes the details of the computational technique and the melt-quench simulation procedure. Section 3 describes the results of this study obtained on the final silica model. Specifically, properties like radial distribution function (RDF), partial pair distribution functions (PPDF), coordination number, bond angle distributions, and neutron structure factor S(Q), rings size distribution calculation, void fraction and void distribution are studied. Apart from the structural analysis, electronic properties are also studied including a detailed analysis of the defect states. Finally, we present a brief conclusion of the work in section 4.

**2. Computational Method:**

In the Born-Oppenheimer molecular-dynamics (BOMD) approach [45], the electronic dynamics is incorporated efficiently by allowing full relaxation of the electrons self-consistently using density functional theory (DFT) every time for a set of ionic configurations. The nuclei positions are fixed here and the Kohn-Sham equations are solved self-consistently to achieve electronic minimization. The electrostatic forces on each atom are calculated by implementing the Hellmann-Feynman theorem. Finally, the atoms are moved (through classical mechanics) by using a specific time step in order to find the new atomic configurations. This process is repeated till convergence is achieved. The Lagrangian for the system is given by the equation (1)

$$L_{BOMD} = E_{K.E.} - E = \frac{1}{2}\sum m_q \dot{r}_q^2 - E[\rho(r), \{r_q\}] \qquad (1)$$

where, $r_q$ and $r$ stand for nuclear and electronic coordinates, while $\rho(r)$ refers to the electron density functional. In the equation, the first term implies the nuclei kinetic energies while the second term implies the total electronic energy obtained by DFT which acts as the potential energy for the nuclei.

## 2.1. Technical aspects of MD simulation

We have performed BOMD simulations on the crystalline polymorph of $SiO_2$ using Vienna *ab-initio* Simulation Package (VASP) [46] in order to obtain the vitreous model. Projector augmented wave (PAW) pseudopotentials in the generalized gradient approximation (GGA) flavour were used for approximating the exchange-correlation potential in VASP [46]. Our starting structure was β-cristobalite having a density of 1.889 gm/cc. Its standard cell contains 24 atoms and we used a cubic simulation cell of side 15 Å (2×2×2 supercell), containing 192 atoms (64 Si and 128 O) that was formally considered our model system. The plane wave cut-off energy was set at 500 eV, the force convergence parameter was set at $10^{-5}$ eV and calculation was done at gamma point only with a time step of 0.5 fs. This cell was heated in a step-by-step procedure as follows.

First, the temperature of the simulation cell was raised to 300 K and the system was allowed to equilibrate for 5 ps. This was followed by subsequently raising the temperature to 1500 K, 2500 K, then to 3500 K, and subsequently to 7000 K and equilibrating the simulation box at each temperature step. Finally, the melted structure was quenched to 300 K followed by equilibration at the same temperature to reduce the interatomic forces and set-in stress to obtain the final configurations. NVT [47] ensemble, under Nosé-Hoover thermostat [48] was used in the whole procedure. The structure was further relaxed using NPT ensemble which helped in minimising the pressure of the final structure to 0.06 kB and attaining a density of 2.24 gm/cc which is in good agreement with the experimental data i.e., 2.203 gm/cc. Fig.1(a) shows the temperature profile of the whole procedure which was carried out in 35 picoseconds. Such kind of melting procedure allowed the system to forget the memory of its initial crystalline configuration while the direct quench process helped in avoiding crystallisation which is important for the intermediate range order. Fig.1(b-d) show the configurations of the system at different temperatures and Fig.1(e) is the polyhedral representation of the quenched system at 300 K.

Apart from this, one more model was generated using the same melt-quench procedure now simulated solely over NPT ensemble. The model obtained via this process exhibited a density of 2.43 gm/cc which had an error of 10.4% as compared to the experimental value i.e., 2.203 gm/cc. Hence, models obtained using the former procedure only are used in further study, which have density of 2.24 gm/cc (error percentage~ 1.6%).

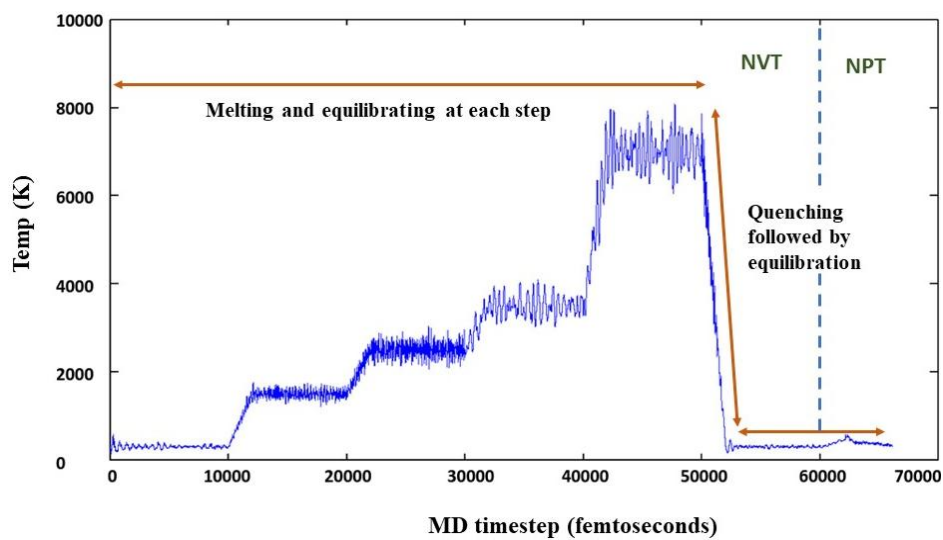

(a)

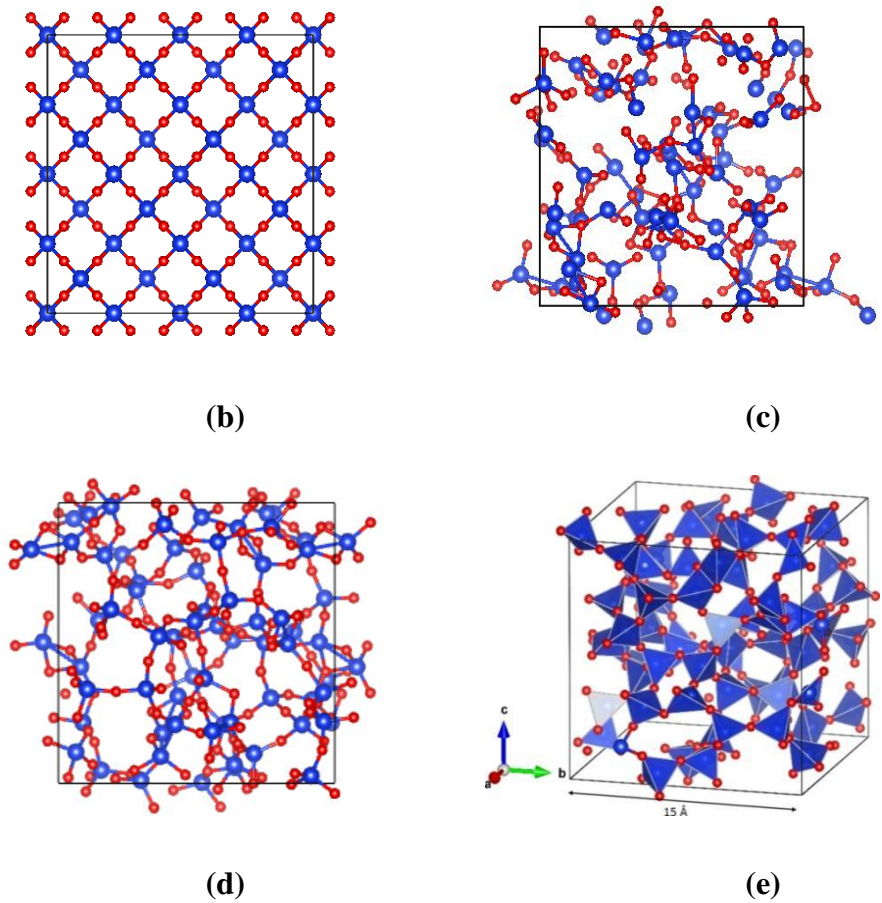

**Fig.1 (a)** Temperature profile of the melt-quench process by MD simulation; Phases of glass transition of $SiO_2$ from crystalline polymorph **(b)** $SiO_2$ crystal at 300 K **(c)** Melted structure at 7000 K **(d)** Quenched structure at 300 K **(e)** Polyhedral representation of vitreous Silica at 300 K (present model)

## 3. Results and discussions

Due to the random arrangement of the atoms in a glassy matrix we need to go to different length scales to characterize it. There is a presence of order in the disordered system in short and intermediate ranges. Since there is absence of any order in the long range, density, and void distribution play an important role in characterizing a glass. These parameters are responsible for the structural properties of the glass. Thus, a detailed study of the properties of one of the obtained structures was done based on these ranges. Further, the current model (AIMD-MQ model) was validated against the existing models like the DFT relaxed continuous random-network model of silica (referred here as the "DFT model") developed by WY Ching et al. [30] (containing 162 atoms in a triclinic cell having dimensions $a$ = 13.72 Å, $b$ = 12.69 Å, and, $c$ = 14.16 Å, $\alpha$ = 91°, $\beta$ = 93.6° and $\gamma$ = 90°), the Potential-free Monte-Carlo model i.e., MC model [15] (2952 atoms in a cubic cell of dimension 36 Å) and other available literature [49-51].

**a)   Short Range Order (SRO ~5 Å)**

The short-range structure is determined by the chemical bonding among the atoms. It usually refers to the arrangement of the atoms over a short range of distance which gives us an insight into the kind of geometry exhibited by the glasses. This arrangement further repeats itself throughout the entire structure. The structural properties that determine the short-range order in glasses are radial distribution function (RDF), bond-angle distribution, coordination number analysis, and structure factor (S(Q)). Experimentally the SRO is determined by the information available from neutron diffraction, NMR, Raman spectroscopy, and X-ray diffraction.

**(i) Radial distribution function (RDF), g(r)**

In a system of particles, the radial distribution gives an insight into the variation of density with the distance from the particle of reference. Fig.2 shows the variation of the total radial distribution function of our system at different temperatures leading to the glass formation. The plot signifies gradual increase in randomness as we go from the crystal polymorph to the amorphous model. The crystalline starting structure has peaks till larger length-scale. Upon melting this structure, all but the first peak disappears implying that no memory of the initial crystalline configuration is retained. Finally, upon quenching and subsequent equilibration glass like peaks were obtained. Unlike the crystalline RDF, the peaks for glassy $SiO_2$ exist only up to the short range and the RDF attains unity in the long-range signifying no long-range order. This agrees with the general understanding of crystal, glass, and melt (liquid) structures and validates the melt-quench simulation protocol adopted here [52]. The total RDF at 300 K for the final glassy system (AIMD-MQ model) is presented in Fig.3(a) and compared with other models. The results were compared with the available experimental [53] and computational data [15,30] and were found to be in good agreement. The first peak at 1.6 Å corresponds to the Si-O bond length while the second peak at 2.6 Å corresponds to the O-O bond length. Subsequently, there is attenuation of the peaks in the RDF of the vitreous model due to the increased randomness in the structure. Hence the RDF approaches unity with the increasing radial distance which is the bulk density. The total RDF plot tends to unity at a radial distance of 4.5 Å. The plots of the partial pair distribution functions (PPDF) of AIMD-MQ model are shown in Fig.3(b). For the Si-O plot, a sharp peak can be seen at 1.6 Å which corresponds to the first neighbour as compared to its crystalline polymorph. But unlike the crystal, after the first sharp peak, there is broadening of the peaks when it comes to the second neighbouring atoms due to the randomness involved. Finally, there is attenuation of the peaks and the PPDF reaches unity. Similar trend is followed by the O-O and Si-Si PPDF plots. These results are in good agreement with the existing literature as well [15,30].

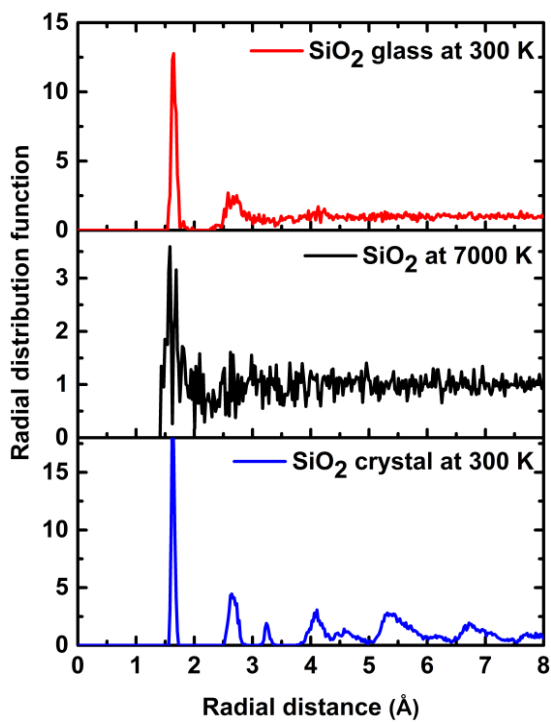

**Fig.2.** Radial distribution function plot for $SiO_2$ crystal at 300 K (in blue), $SiO_2$ melt at 7000 K (in black), and $SiO_2$ glass at 300 K (in red).

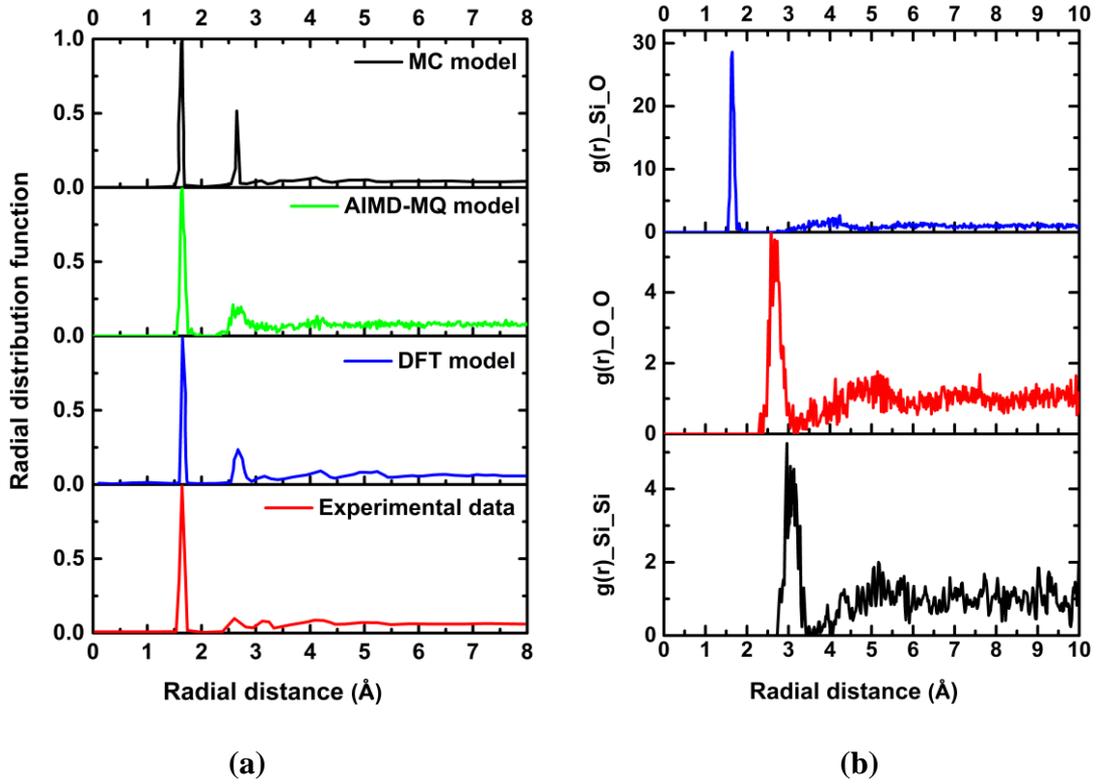

**(a)**                           **(b)**

**Fig.3 (a)** Radial distribution function plot for the AIMD-MQ model (in green) compared with the MC model (in black), DFT model (in blue), and experimental data (in red); **(b)** Partial pair distribution functions of Si-O (in blue), O-O (in red), and Si-Si (in black) for AIMD-MQ model

**(ii) Bond angle distribution**

Fig.4(a) shows the intra-tetrahedral distribution of O-Si-O bond angles [54] for amorphous $SiO_2$ at 300 K (five configurations of the present model were considered for the calculation). The angle varies between 80º to 120º which is the usual nature exhibited by a random network structure. The distribution peaks at around 108º which is contributed by the directional bonds that are very strong by nature, constituting the glassy matrix. Similar distributions are observed for the DFT model [30] whereas the peak formed in the MC model [15] is sharper because the structure was not force relaxed. Fig.4(b) shows the inter-tetrahedral Si-O-Si bond angle distribution. The distribution has a peak around 138º which is smaller than the experimental value where this angle varies between 120º and 180º and peaks around 145º [55]. Fig.4(c) shows the plot of the dihedral angle distribution in our structure. These results are in good agreement with the MC and DFT studies [15,30].

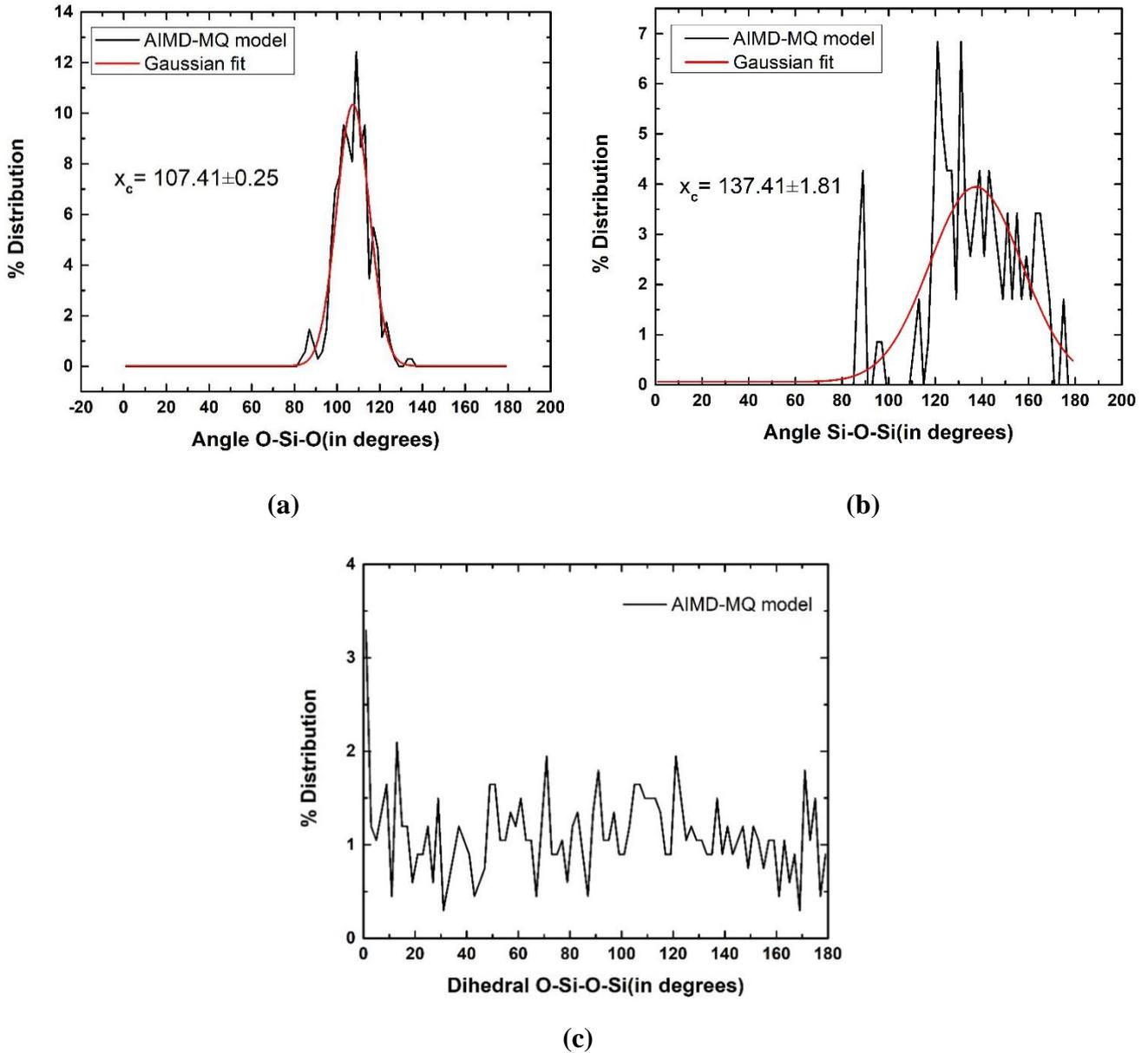

**Fig.4.** (a) intra-tetrahedral bond angle, (b) inter-tetrahedral bond angle, (c) Dihedral angle distributions

**(iii) Coordination number**

Coordination number (CN) analysis for both Si and O atoms was done. Five configurations were used and the average coordination numbers for Si and O atoms were found to be 3.9 and 1.9, respectively. For an ideal structure without any coordination defects, the average CN value is 4 for Si atom and 2 for O atom. But realistic models of glass should contain coordination defects since they have been observed experimentally as well. Their percentages vary depending upon the preparation conditions [56]. The details of the analysis for the configurations around both the atoms for AIMD-MQ model along with other models are shown in Table 1.

Due to the variation in the coordination numbers, the average Si-O bond length in each tetrahedral unit varied as well. Out of the 64 Si atoms present in our system (AIMD-MQ), 55 atoms were tetrahedrally coordinated containing both corner sharing (49) and edge sharing (6) tetrahedra. An ideal glassy structure contains only corner sharing tetrahedra but experimentally observed structures contain some defects which are similar to the edge sharing tetrahedra present in our system. Table 2 gives the distribution of the average bond lengths for different coordination numbers of Si in our system.

**Table 1. Coordination number analysis for configurations around Si and O atom.**

| CN of Si | % Distribution in AIMD-MQ model | % Distribution in DFT model | % Distribution in MC model | CN of O | % Distribution in AIMD-MQ model | % Distribution in DFT model | % Distribution in MC model |
|---|---|---|---|---|---|---|---|
| 1 | 0 | 0 | 1.01 | 0 | 1.56 | 0 | 0 |
| 2 | 3.12 | 0 | 13.6 | 1 | 3.91 | 0 | 40.8 |
| 3 | 9.4 | 0 | 2.8 | 2 | 94.53 | 100 | 58.7 |
| 4 | 86 | 100 | 82.4 | 3 | 0 | 0 | 0.3 |
| 5 | 1.6 | 0 | 0 | 4 | 0 | 0 | 0 |
| Average CN | 3.86 ±0.05 | 4 | 3.66 | Average CN | 1.93 ±0.02 | 2 | 1.59 |

**Table 2. Distribution of the Si-O bond lengths in the AIMD-MQ model.**

| CN of Si | No. of polyhedral configurations | % Distribution | Average bond lengths of Si-O bonds (Å) |
|---|---|---|---|
| 1 | 0 | 0 | 0 |
| 2 | 2 | 3.12 | 1.73 |
| 3 | 6 | 9.4 | 1.75 |
| 4 | 55 | 86 | 1.64 |
| 5 | 1 | 1.6 | 1.73 |

**(iv) Calculated neutron structure factor S(Q)**

Fig.5 shows the neutron structure factor calculated from the Fourier transform of g(r) for the present model, which is compared with the experimental data and other models. The peaks in S(Q) agree well with that of experimental data as well as with other models. We can also find the presence of FSDP (shown with dotted vertical line in the figure) [35-37] at ~1.49±0.02 Å$^{-1}$ which is in good agreement with experimental data [57], which signifies the presence of MRO in the glassy structures. In the DFT model, an extra peak can be seen at 0.7 Å$^{-1}$. Such a peak is undesirable as they represent periodicity at length scales greater than medium-range which is not true for glasses and are an artefact due to the finite size of the periodic system in the calculations. Comparison of Full Width at Half Maximum (FWHM) of the FSDP for these models is shown in the Table 3.

From earlier simulation studies [58,59] on the dependence of FSDP on size of the structure, it was found that the intensity and the FWHM of FSDP depend on the system size; however, the position of FSDP can even be reproduced using small systems with around 200 atoms. So, from the present AIMD model, only the Q position of the FSDP can be ascertained and other properties like intensity and FWHM of FSDP may change if a larger cell size is generated, a study not undertaken at present due to computational constraints.

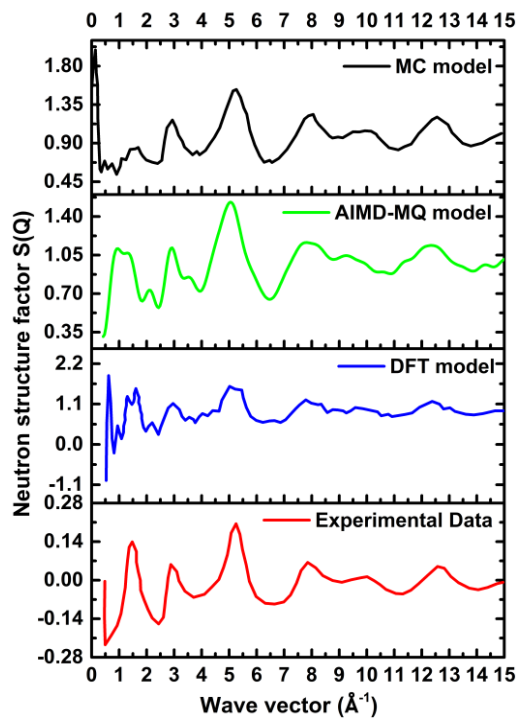

**Fig.5.** Neutron structure factor S(Q) plot for the AIMD-MQ model (in green) compared with MC model (in black), DFT model (in blue), and Experimental data (in red).

**Table 3. FSDP positions and corresponding FWHMs for various models of a-SiO$_2$.**

| Structural model | Position of FSDP (in Å$^{-1}$) | FWHM (in Å$^{-1}$) |
|---|---|---|
| MC model | 1.62 | 0.64 |
| AIMD-MQ model | 1.49 | 0.51 |
| DFT model | 1.53 | 0.55 |
| Experimental | 1.53 | 0.53 |

**b)    Medium Range Order (MRO ~ 5 Å to 10 Å)**

It is difficult to characterize a glass at this range due to the randomness involved in the structure. However, experimentally it is found that the FSDP [60] occurring in the total structure factor of a glassy system signifies the existence of the medium range order. Our system is also in good agreement with this study (Fig.4). Apart from this property, no other experimental technique has been designed to quantify this order in vitreous systems; while computationally Ring size distribution gives us an insight into the MRO in glasses. A ring is defined as the shortest closed path formed by the nodes and the links; where nodes refer to the number of atoms while the link signifies the bonds between the nodes. There are two ways of determining the size of rings. First, we can count the total number of nodes/atoms present in the ring i.e., a ring containing $N$ nodes is an $N$-membered ring. Second, we can count the number of network-forming nodes i.e., for our system, Si is known as the network former since this has the highest coordination. So, in the second method, an $N$-membered ring contains $2N$ number of nodes. Since we would be following Guttman's shortest path criterion [61] we prefer the first convention for numbering the rings. According to Guttman a ring is the shortest closed path starting from a node and coming back to the same from its nearest neighbouring atoms. By using the R.I.N.G.S software the distribution of various types of rings ($N$-membered rings) can be studied in the structure [62].

Vitreous SiO$_2$ contains -Si-O-Si-O type of rings. In the distribution of various ring sizes in a silica glass network, it is found that the 12-membered rings appear the most throughout the glassy matrix. We observe a similar type of distribution in our study as well. Fig.6 shows a ball and stick representation of types of rings present in the system; Four member rings seen arise from the edge shared tetrahedral defects that are seen in the models. Fig.7 shows the ring size distribution in our model where there is a peak at the 12-node rings. This is in good agreement with studies on the distribution of the ring size in vitreous Silica by Roux et al. and Kohara et al. [62,63].

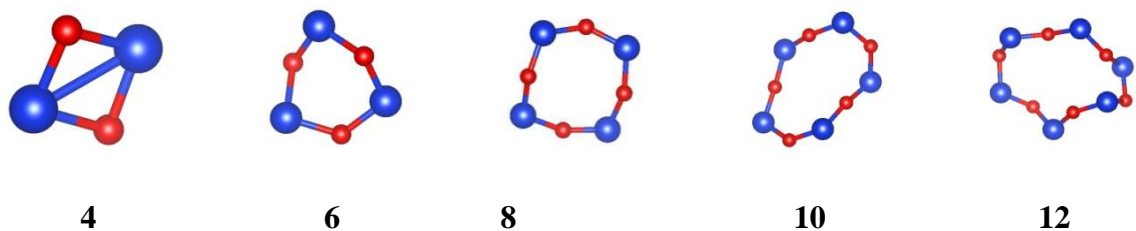

**Fig.6.** Different membered rings present in the glassy matrix.

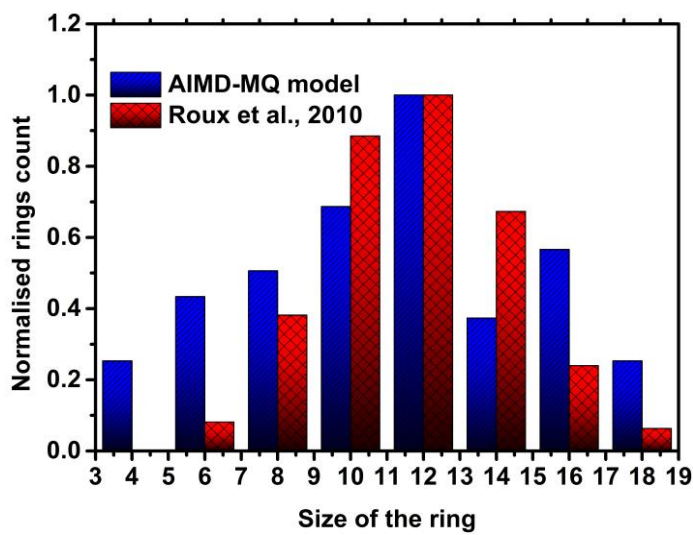

**Fig.7.** Ring size distribution in the present model as compared to study by Roux et al. [62,63]

## (c) Long-Range structure (> 10 Å)

In the case of vitreous systems there is the absence of any order in the long range because of the increased randomness in the structure. So, to characterize a glass in this range we resort to some macroscopic properties like density and void distribution. The calculated values of these two properties of the AIMD-MQ model obtained computationally were compared with the available experimental data. The density of our structure was found to be 2.24 gm/cc while the experimental density value for vitreous $SiO_2$ is 2.203 gm/cc [64]. Hence, our result is in good agreement with the experimental data with an error of 1.6 %. Similarly, from the void distribution analysis, the value of the void fraction was calculated to be 0.29. MOF explorer software [65] was used for this study. Table 4 shows a detailed comparison of the void fractions of different models. The AIMD-MQ model's result is in good agreement with the available theoretical data but shows a variation when compared with the experimental data. This deviation can be accounted for by the usage of Van der Waal's radius of the atoms in theoretical models which may not resemble the experimental model. Experimentally, the void fraction is calculated using the positron probe i.e., positron annihilation lifetime spectroscopy [66]. Fig.8 shows the void distribution curve for the AIMD-MQ model, DFT model, and MC model. The result of the AIMD-MQ model is in good agreement with the rest of the models. The first peak in the model accounts for that volume present in the system, which can be occupied by no atom known as the excluded volume whereas the last peak at about 4 Å is the artefact of the small size of the structure. Apart from this, the size distribution of voids has a peak at a radius of 0.9 Å, which compares well with other computational models (MC and DFT model [15,30]) discussed in the study. Also, the distribution compares well with other studies of void structure of silica glass [67-69] which shows a near Gaussian distribution of void size with peak at radius ~0.8-0.9 Å.

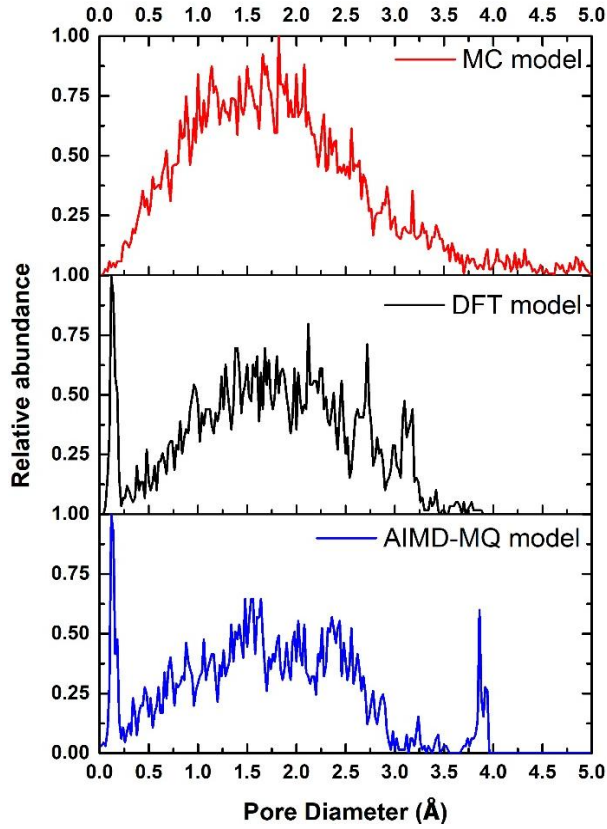

**Fig.8.** Void distribution curve of the present model (in blue) compared with the DFT model (in black) and MC model (in red).

**Table 4. Comparison of density and void fraction values of the present model with experimental and theoretical data.**

| Structural Model | Density(gm/cc) | Void fraction |
|---|---|---|
| DFT Model | 2.19 | 0.31 |
| MC Model | 2.07 | 0.33 |
| Experimental data | 2.20 | 0.18 |
| AIMD-MQ Model | 2.24 | 0.29 |

**(d) Electronic structure**

In this section, we have studied the electronic density of states (DOS). The total DOS of vitreous $SiO_2$ at 0 K is shown in Fig.9. According to previous computational studies, the band gap for a perfectly coordinated model of amorphous $SiO_2$ is calculated to be ~5 eV [30]. For the AIMD model, the band gap is found to be around 4.9 eV which agrees well with the DFT studies [30]. Apart from this, a few defect peaks can also be seen in the DOS plot.

Fig.11(a) shows the partial density of states (PDOS) plot for the Si and O atoms that are responsible for the formation of the dangling bonds. It is observed that these defects give rise to the deep and donor levels in the band gap region. Fig.11(b) is the PDOS plot for the Si and O atoms constituting another coordination defect i.e., edge sharing tetrahedral units. In this case there are no defect peaks in the band gap region and mostly the contribution to the DOS in the valence band.

In addition, there are also other types of defects like vacancies and bridging oxygen defects. Type I bridging oxygen connects two Si atoms, while the type II bridging oxygen connects two oxygen atoms. We have quantified the effects of such defects in the obtained glassy matrix as well, by creating them in the models and studying the changes in the DOS. Fig.10 shows a pictorial representation of such defects. Fig.12 shows the density of states plot for the AIMD model after creating the O and Si vacancies and the two types of the bridging oxygen defects. It is seen that deep levels are created in all the three DOS plots. These peaks signify the defect/trap states. These defect states affect the optical properties of a material by introducing additional energy levels within the band gap leading to changes in its optical absorption and emission properties. Apart from this, acceptor levels have formed in all the three cases while donor levels formed in case of Si vacancy are negligible compared to those formed in the O vacancy and the bridging oxygen DOS plots.

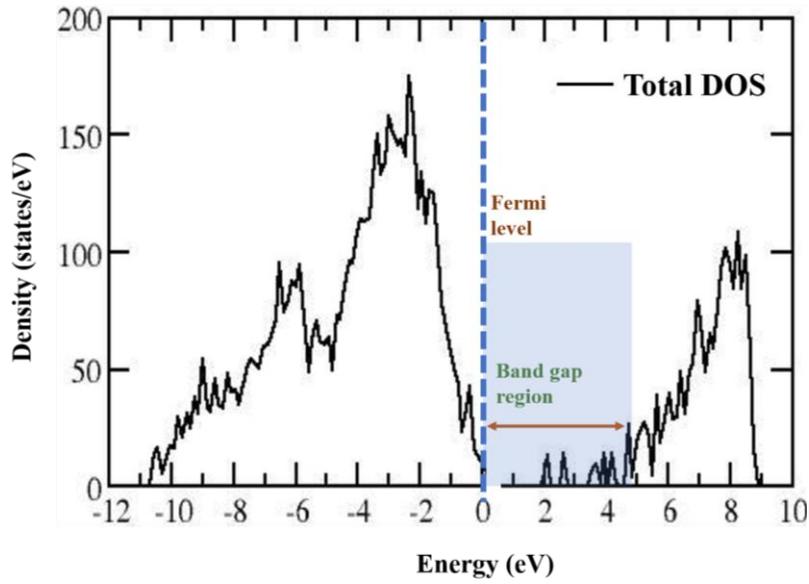

**Fig. 9.** Total electronic density of states of vitreous Silica, the peaks in band gap region are the defects/trapping states

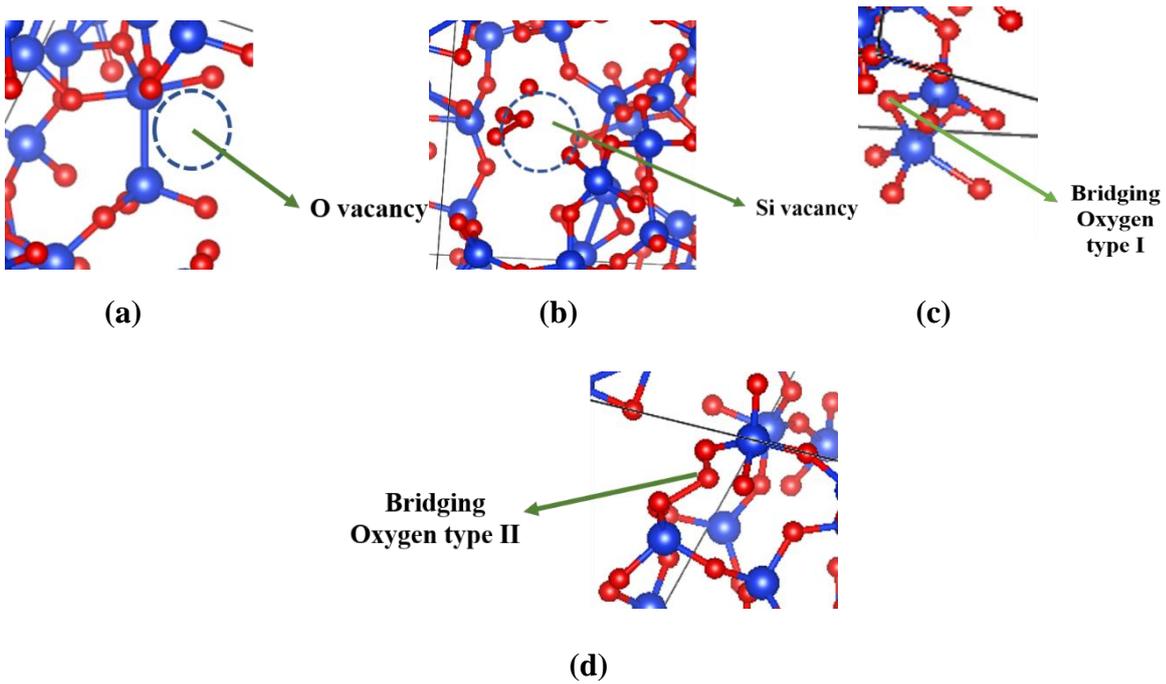

**Fig.10.** Defects found in a glassy matrix (a) Oxygen vacancy, (b) Silicon vacancy, (c) Bridging Oxygen type I, and (d) Bridging Oxygen type II

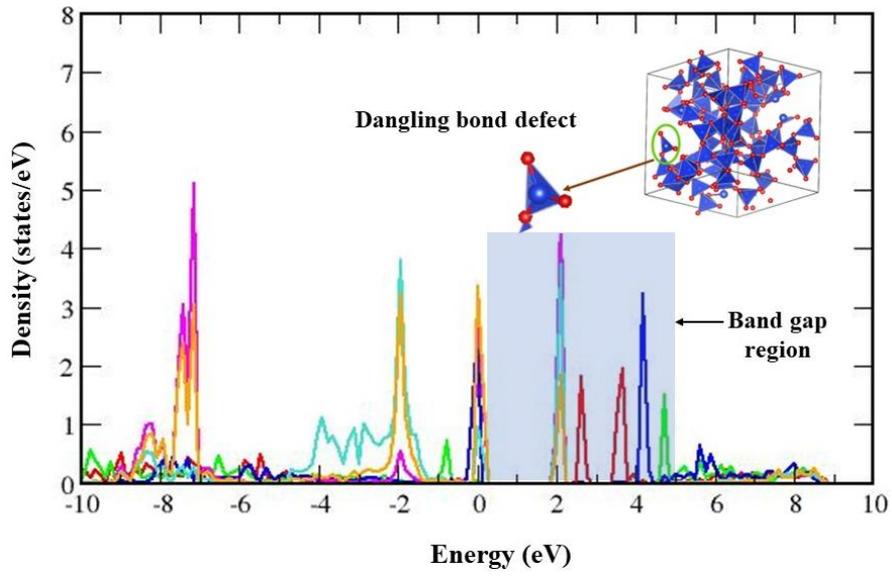

(a)

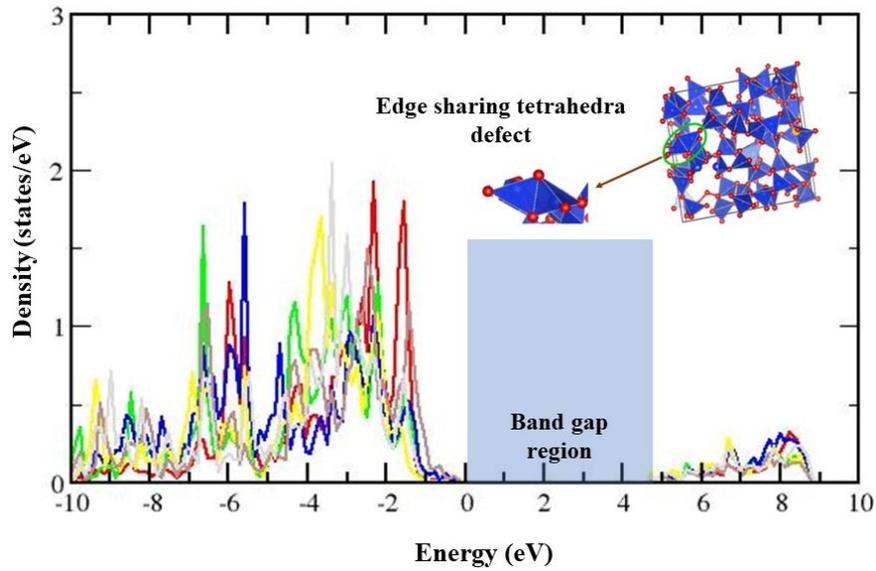

(b)

**Fig.11.** Partial density of states plots of (a) atoms responsible for the formation of dangling bonds (b) atoms responsible for the formation of edge sharing connection of tetrahedral units

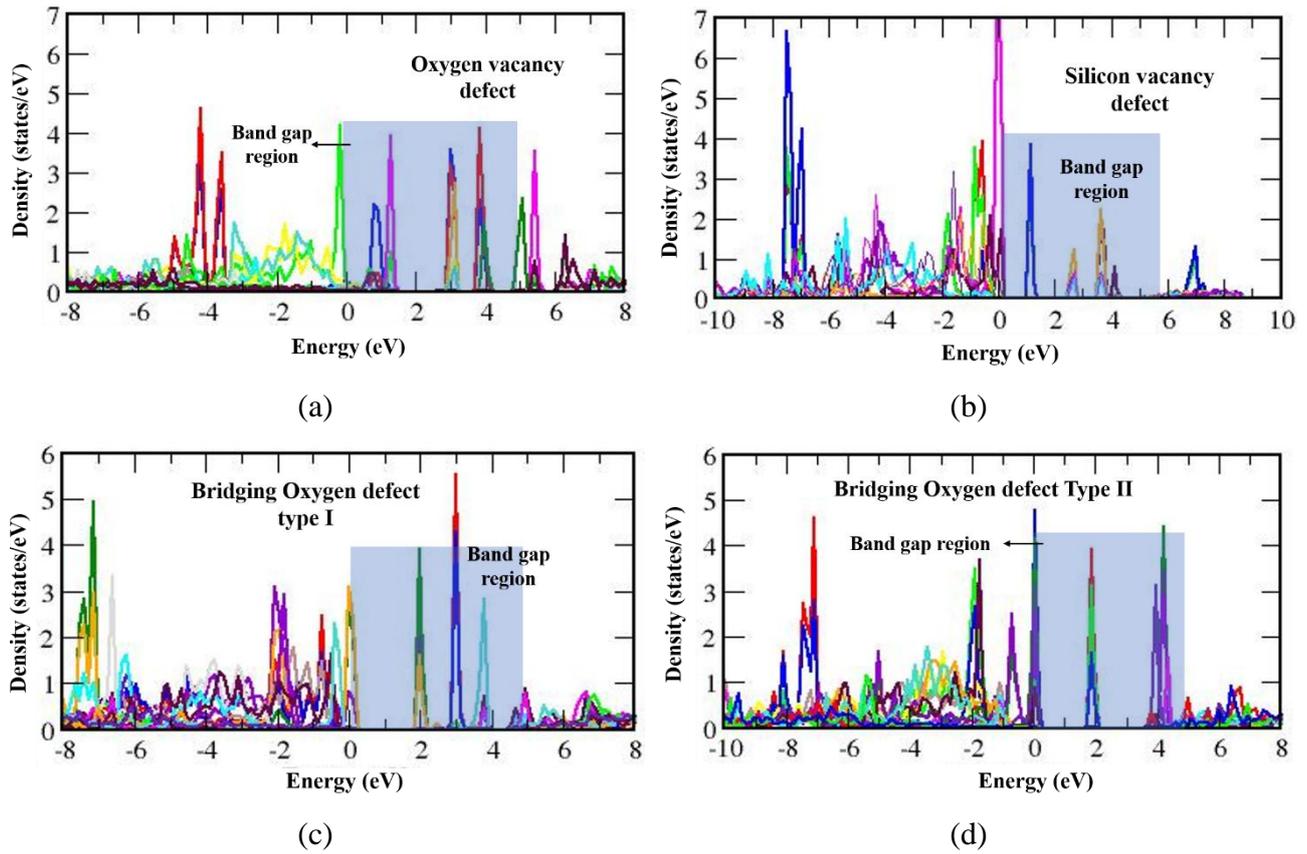

**Fig.12.** Partial density of states plot for (a) Oxygen vacancy, (b) Silicon vacancy, (c) Bridging Oxygen type I, and (d) Bridging Oxygen type II defects created in the system.

## 4. Conclusion

We have presented models of vitreous silica prepared by simulating the melt-quench technique through *ab-initio* molecular dynamics simulations. To validate the model in the short-range, properties like radial distribution function, coordination number, and bond angle distributions were studied. To characterize the system in the medium range, neutron structure factor was obtained and a detailed analysis of ring distribution in the system was done. Finally, in the long-range, properties like void fraction and density were calculated. All the above properties were compared with the existing atomistic models and experimental data and were found to be in good agreement with the same. Apart from the structural properties, electronic properties were studied as well. From this study, it was concluded that the dangling bonds present in the glassy network, contributed to the trapping states in electronic DOS in the band gap. The other defects in the system such as the edge-sharing tetrahedra showed no contribution to the electronic states in the band gap. In addition, bridging Oxygen defect as well as Oxygen and Silicon vacancies were created and their effects on the system were thoroughly studied. The *ab-initio* melt-quench simulation of glasses has to necessarily tackle the problems associated with high quench rates and system size limits. By choosing an appropriate quench rate and a subsequent relaxation using NPT ensemble and conjugate gradient method, the present study has shown that it is possible to make glassy models that agree with experimental data not only on the short-range scales but also up to the medium-range scales by reproducing the desired rings distribution and FSDP in S(Q) thereby exhibiting MRO.

**Acknowledgement**

Sruti S Jena is thankful to Dr. Gurpreet Kaur, Materials Science Group, Indira Gandhi Centre for Atomic Research (IGCAR), for help with the VASP software and useful discussions.

**Data Availability:** The raw/processed data required to reproduce these findings cannot be shared at this time as the data also forms part of an ongoing study.

**Statement of Contribution**

1. Sruti Sangeeta Jena- Computational work, data analysis and interpretation and article writing.
2. Shakti Singh- Data analysis and interpretation and useful discussions.
3. Sharat Chandra- Problem definition, guidance and useful discussions.